\documentclass{PoS}
\usepackage{amsmath}
\newcommand{\red}[1]{\textcolor{red}{#1}}
\newcommand{\blue}[1]{\textcolor{blue}{#1}}

\title{
\begin{picture}(0,0)(0,0)%
   \put(350,55){\makebox(0,0)[l]{\textnormal{\normalsize 
   J-PARC-TH-0157 }}}%
   \end{picture}%
  Exploring non-abelian gauge theory with energy-momentum tensor; stress, thermodynamics and correlations}

\ShortTitle{Exploring non-abelian gauge theory with energy-momentum tensor}

\author{\speaker{Masakiyo Kitazawa} \\ %\thanks{A footnote may follow.}\\
  Department of Physics, Osaka University, Toyonaka, Osaka 560-0043, Japan\\
  J-PARC Branch, KEK Theory Center, Institute of Particle and Nuclear Studies, KEK, 203-1, Shirakata, Tokai, Ibaraki 319-1106, Japan\\
        E-mail: \email{kitazawa@phys.sci.osaka-u.ac.jp}}

\abstract{
  We perform various lattice numerical analyses with the energy-momentum
  tensor (EMT) defined through the gradient flow.
  We explore the spatial distribution of the stress tensor
  in static quark-anti-quark systems and thermodynamic quantities
  at nonzero temperature, as well as the correlation functions of EMT.
  The stress tensor distribution is also studied in the Abelian-Higgs model,
  which is compared with the lattice result.
}

\FullConference{XIII Quark Confinement and the Hadron Spectrum - Confinement2018\\
		31 July - 6 August 2018\\
		Maynooth University, Ireland}

\begin{document}

\section{Introduction}

The energy-momentum tensor (EMT) 
\begin{align}
  T^{\mu\nu}(x)
\end{align}
is one of the most fundamental observables in physics.
It's temporal and spatial components are related to 
important quantities in physics, i.e.
energy density $\varepsilon$, momentum density $P^i$,
and the stress tensor $\sigma_{ij}(x)$, as 
\begin{align}
  \varepsilon(x) = T^{00}(x), \quad
  P^i(x)=T^{0i}(x), \quad
  \sigma_{ij}(x) = - T^{ij}(x) \qquad (i,j=1,2,3).
  \label{eq:epsigma}
\end{align}
Among these observables, 
the stress tensor $\sigma_{ij}(x)$ is a particularly interesting
quantity because it represents the distortion of the field
which mediates the force between charges.
The direct analysis of the stress tensor in various systems in QCD,
such as static-quark systems and hadrons, 
will provide us with deeper understanding
on these systems based on microscopic points of view
in a gauge invariant manner.
In a thermal system at nonzero temperature,
$\sigma_{ij}(x)$ is given by a diagonal matrix representing
the pressure, which is a basic information on thermodynamics.

Recently, considerable developments have been made in numerical
analyses of EMT in lattice gauge theory~\cite{Suzuki:2016ytc}.
In particular, it was found~\cite{Suzuki:2013gza} that the analysis of
EMT on the lattice can be performed successfully with the use of
the gradient flow~\cite{Luscher:2010iy,Narayanan:2006rf,Luscher:2011bx}.
Because EMT is a fundamental observable in physics,
its analyses on the lattice will provide us with new insights
into QCD and non-Abelian gauge theories.

In this proceeding,
after introducing the EMT operator on the lattice constructed from 
the gradient flow in Sec.~\ref{sec:EMT},
we discuss recent applications of the EMT operator to the analysis of
various quantities, thermodynamics (Sec.~\ref{sec:therm}),
EMT correlation functions (Sec.~\ref{sec:cor}), and
the stress distribution in the $Q\bar{Q}$ system (Sec.~\ref{sec:qqb}).

\section{Energy-momentum tensor and gradient flow}
\label{sec:EMT}

Let us first see the construction of EMT
using the gradient flow~\cite{Suzuki:2013gza}.
The gradient flow for the YM theory is a 
continuous transformation of the gauge field $A_\mu(x)$ defined by the 
differential equation~\cite{Luscher:2010iy,Narayanan:2006rf,Luscher:2011bx}
\begin{align}
\frac{d A_\mu(t,x)}{dt} = - g_0^2 
\frac{ \delta S_{\rm YM}(t)}{ \delta A_\mu(t,x) }
= D_\nu G_{\nu\mu}(t,x) ,
\label{eq:GF}
\end{align}
with the Yang-Mills action $S_{\rm YM}(t)$ composed of the field
$A_\mu(t,x)$ at nonzero flow time $t$.
%Color indices are suppressed for simplicity.
The initial condition at $t=0$ is taken for the 
conventional gauge field; $A_\mu(0,x)=A_\mu(x)$.
The flow time $t$, which controls the magnitude of transformation,
has a dimension of inverse mass squared.
At the tree level, Eq.~(\ref{eq:GF}) is written as 
\begin{align}
  \frac{d A_\mu}{dt} = \partial_\nu \partial_\nu A_\mu
  + {\rm (gauge ~ dependent ~ term)}.
\label{eq:diffusion}
\end{align}
Neglecting the gauge dependent term, Eq.~(\ref{eq:diffusion})
is the diffusion equation in four-dimensional space.
Therefore, the gradient flow for positive $t$ acts as 
a cooling of the gauge field with smearing radius 
$\sqrt{8t}$.

In the present study, we use the gradient flow to introduce
the EMT operator using the small flow time expansion 
(SFTE) \cite{Luscher:2011bx,Suzuki:2013gza}.
The SFTE asserts that
a composite operator $\tilde{O}(t,x)$ 
composed of the field $A_\mu(t,x)$ at $t>0$
is represented in terms of 
the operators in the original gauge theory as 
\begin{align}
  \tilde{O}(t,x) \xrightarrow[t\to0]{} \sum_i c_i(t) O_i^{\rm R}(x) ,
  \label{eq:SFTE}
\end{align}
in the small $t$ limit, 
where $O_i^{\rm R}(x)$ on the right-hand side are renormalized 
operators in the original gauge theory at $t=0$ with the subscript $i$
denoting different operators.

In order to construct EMT using Eq.~(\ref{eq:SFTE}),
we expand the following operators via the SFTE;
\begin{align}
  U_{\mu\nu}(t,x) &= G_{\mu\rho}^a (t,x)G_{\nu\rho}^a (t,x)
  -\frac14 \delta_{\mu\nu}G_{\rho\sigma}^a(t,x)G_{\rho\sigma}^a(t,x), 
  \label{eq:U}
  \\
  E(t,x) &= \frac14 G_{\mu\nu}^a(t,x)G_{\mu\nu}^a(t,x).
  \label{eq:E}
\end{align}
The SFTEs of Eqs.~(\ref{eq:U}) and (\ref{eq:E}) are given by 
\begin{align}
   U_{\mu\nu}(t,x)
   &=\alpha_U(t)\left[
   T_{\mu\nu}^R(x)-\frac14 \delta_{\mu\nu}T_{\rho\rho}^R(x)\right]
   +O(t),
\label{eq:(2)}\\
   E(t,x)
   &=\left\langle E(t,x)\right\rangle_0
   +\alpha_E(t)T_{\rho\rho}^R(x)
   +O(t),
\label{eq:(3)}
\end{align}
where $\langle\cdot\rangle_0$ denotes vacuum expectation value and 
$T_{\mu\nu}^R(x)$ is the correctly renormalized EMT.
Abbreviated are 
the contributions from the operators of dimension~$6$ or higher, 
which are proportional to powers of $t$ because of dimensional 
reasons and suppressed for small $t$.

Combining Eqs.~(\ref{eq:(2)}) and~(\ref{eq:(3)}), we obtain
\begin{align}
   T_{\mu\nu}^R(x)
   =\lim_{t\to0} T_{\mu\nu}(t,x);
   \quad
   T_{\mu\nu}(t,x)
   =c_1(t) U_{\mu\nu}(t,x)
   +c_2(t) \frac{\delta_{\mu\nu}}{4}
   \left[E(t,x)-\left\langle E(t,x)\right\rangle_0 \right].
\label{eq:T^R}
\end{align}
The coefficients $c_1(t)$ and $c_2(t)$ are calculated
perturbatively up to one- and two-loop orders, respectively,
in Ref.~\cite{Suzuki:2013gza}%
\footnote{
  The perturbative analyses of $c_1(t)$ and $c_2(t)$ are
  recently extended to one more higher order;
  see Refs.~\cite{Harlander:2018zpi,Iritani:2018idk}.
}.
We use these coefficients in the following analysis.

The concept of the gradient flow and 
the construction of EMT via the SFTE can also be extended to full
QCD with fermions~\cite{Luscher:2013cpa,Makino:2014taa,
  Taniguchi:2016ofw,Taniguchi:2016tjc}.
In this case, one needs five operators for the SFTE of EMT;
in addition to Eqs.~(\ref{eq:U}) and (\ref{eq:E}), there are
three operators including fermions at dimension $4$.
The coefficients in the SFTE in this case is calculated in
Ref.~\cite{Makino:2014taa} (see also Ref.~\cite{Harlander:2018zpi}).

From Eq.~(\ref{eq:T^R}), one can obtain $T^R_{\mu\nu}(x)$
in the numerical simulation of lattice gauge theory by 
the following procedure:
\begin{enumerate}
\item
  Generate gauge configurations at $t=0$ 
  with a standard algorithm.
\item
  Obtain the flowed gauge field for $t>0$ by 
  numerically solving the flow equation (\ref{eq:GF}).
\item
  Analyze $U_{\mu\nu}(t,x)$ and~$E(t,x)$ on the flowed field at each $t$,
  and determine $T_{\mu\nu}(t,x)$ in Eq.~(\ref{eq:T^R}).
  Then construct the expectation value or correlation functions of
  $T_{\mu\nu}(t,x)$.
\item
  Carry out the double extrapolation to $(t,a)=(0,0)$
  where $a$ is the lattice spacing.
\end{enumerate}

In the last step for the double extrapolation, 
the analysis has to be performed in the parameter range
satisfying $a\lesssim\sqrt{2t}\lesssim R$, where
$R$~is an infrared cutoff scale such as~$\Lambda_{\text{QCD}}^{-1}$,
or the shortest length relevant for the problem such as
$T^{-1}=N_\tau a$ for temperature $T$ and distances between operators.
The condition $a\lesssim\sqrt{2t}$ is necessary to suppress 
the finite $a$ correction which diverges for $t\to0$.

\section{Thermodynamics}
\label{sec:therm}

\begin{figure}[t]
 \centering
 \includegraphics[width=0.49\textwidth,clip]{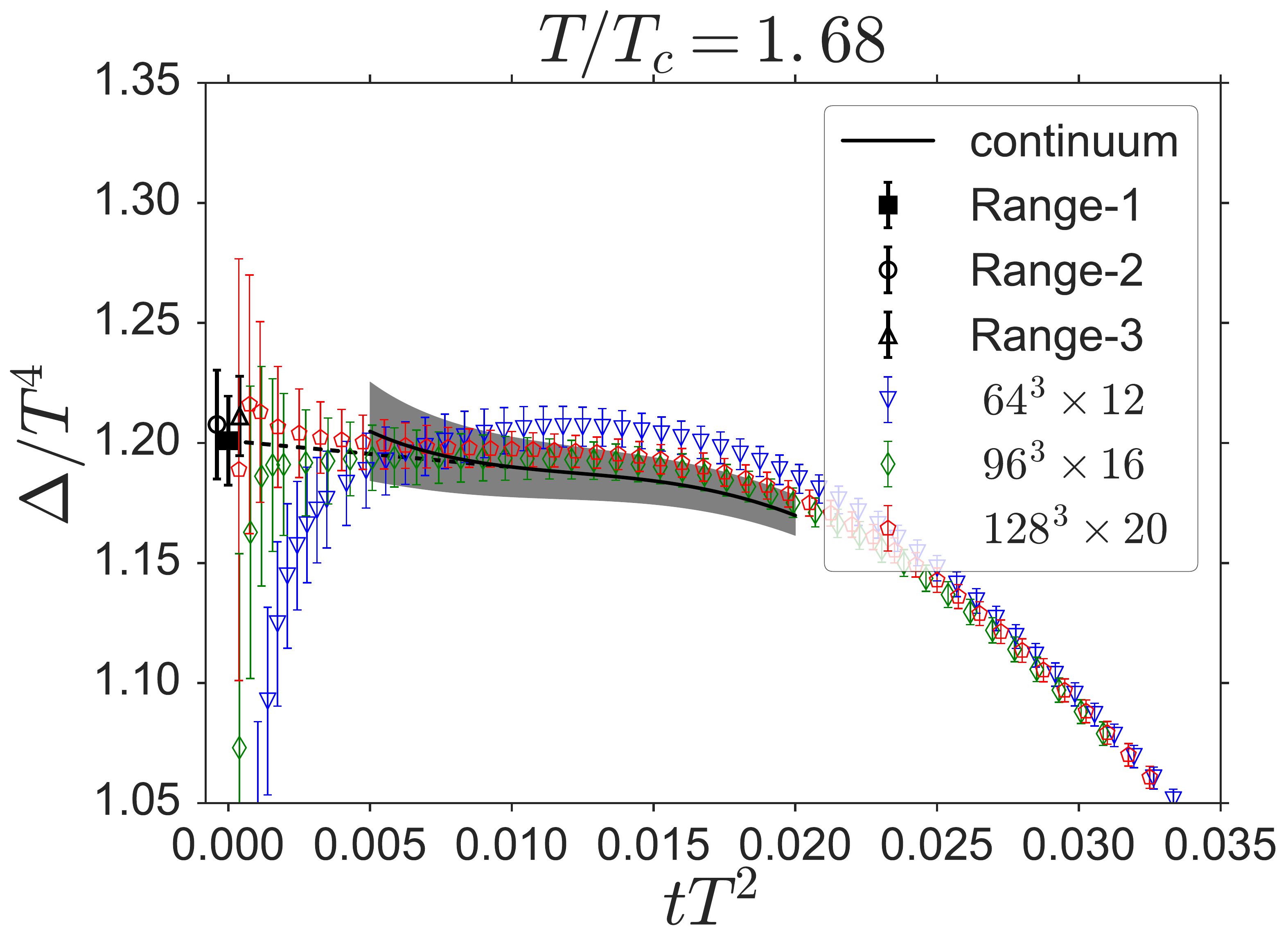}
 \includegraphics[width=0.49\textwidth,clip]{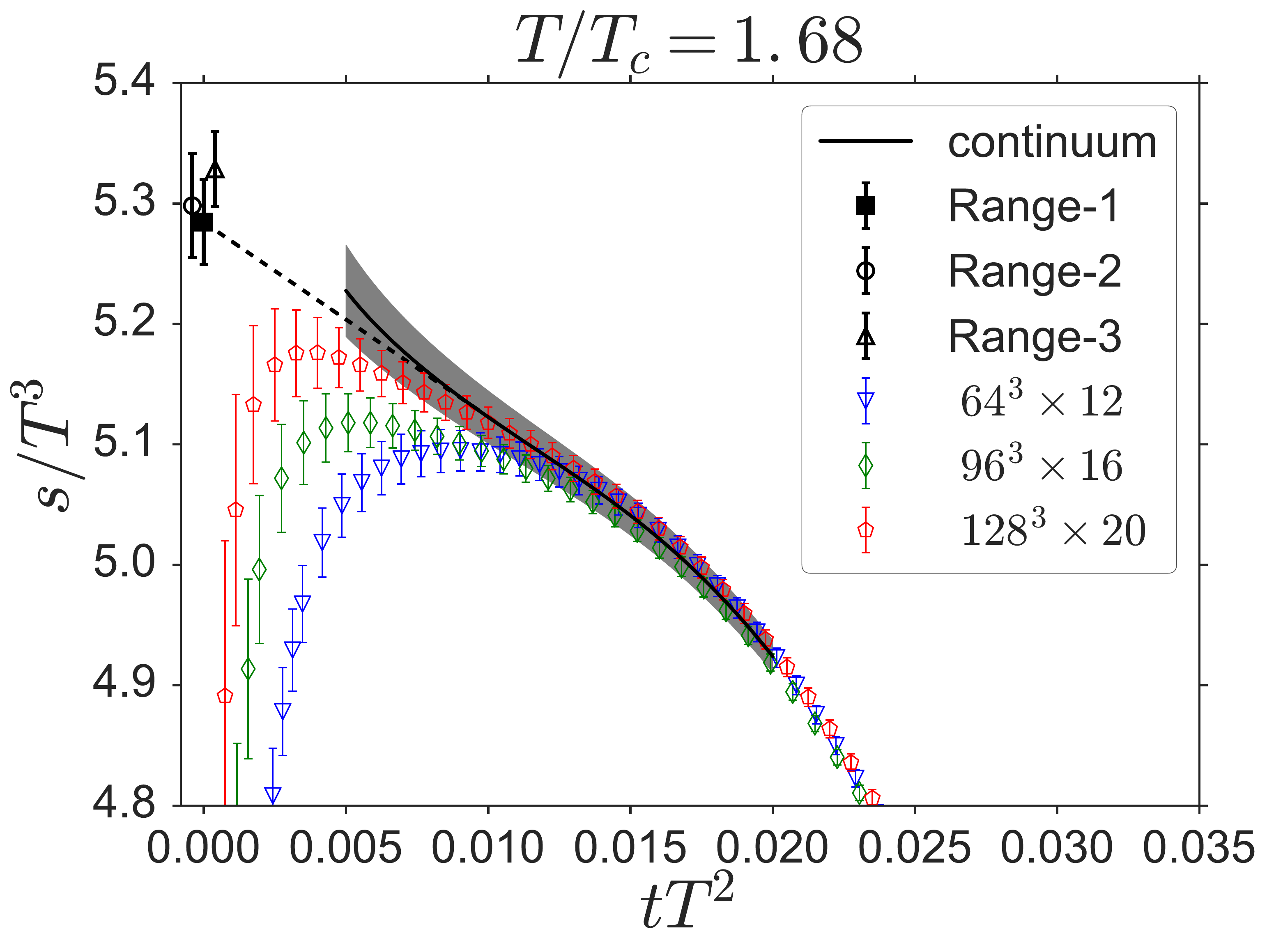}
 \caption{
   Behaviors of $\Delta(t)$ and $s(t)$ as functions of
   the flow time $t$ at $T/T_c=1.68$ in SU(3) YM theory~\cite{Kitazawa:2016dsl}.
   The points denote numerical results obtained on the lattice
   with three different lattice spacings.
   The black line denotes the result of the continuum
   extrapolation with fixed $t$.
   The values around $tT^2=0$ show the results of $t\to0$ extrapolation
   obtained with three different fitting ranges.}
 \label{fig:therm168}
\end{figure}

Now let us apply the EMT operator defined in
the previous section to the analysis of thermodynamic
quantities in SU(3) YM theory~\cite{Asakawa:2013laa,Kitazawa:2016dsl}.
In the following, we consider 
$\Delta=e-3p$ and the entropy density $s=(e+p)/T$ given by 
linear combinations of the energy density $e$ and pressure $p$.

In Fig.~\ref{fig:therm168}, we plot
\begin{align}
  \Delta(t) = - \sum_{i=1}^4 \langle T_{ii}(t,x) \rangle, \quad
  s(t) = \frac1T\big( - \langle T_{44}(t,x) \rangle
  + \frac13 \sum_{i=1}^3 \langle T_{ii}(t,x) \rangle \big),
\end{align}
as functions of $t$ at $T/T_c=1.68$ obtained on the lattices with three
different lattice spacings~\cite{Kitazawa:2016dsl}.
To take the double extrapolation $(t,a)\to(0,0)$ from these results,
we first carry out the continuum extrapolation for each $t$.
The result of this extrapolation is plotted by the black line
with errors shown by the shaded region.
We then take the $t\to0$ limit using this continuum
extrapolated result with three different
fitting ranges of $t$; Range-1: $0.01<tT^2<0.015$,
Range-2: $0.005<tT^2<0.015$, Range-3: $0.01<tT^2<0.02$.
The extrapolated values with these ranges are shown in the figure
around $tT^2=0$.
Their difference is taken into account in the systematic error
in the final result.

\begin{figure}[t]
 \centering
 \includegraphics[width=0.49\textwidth,clip]{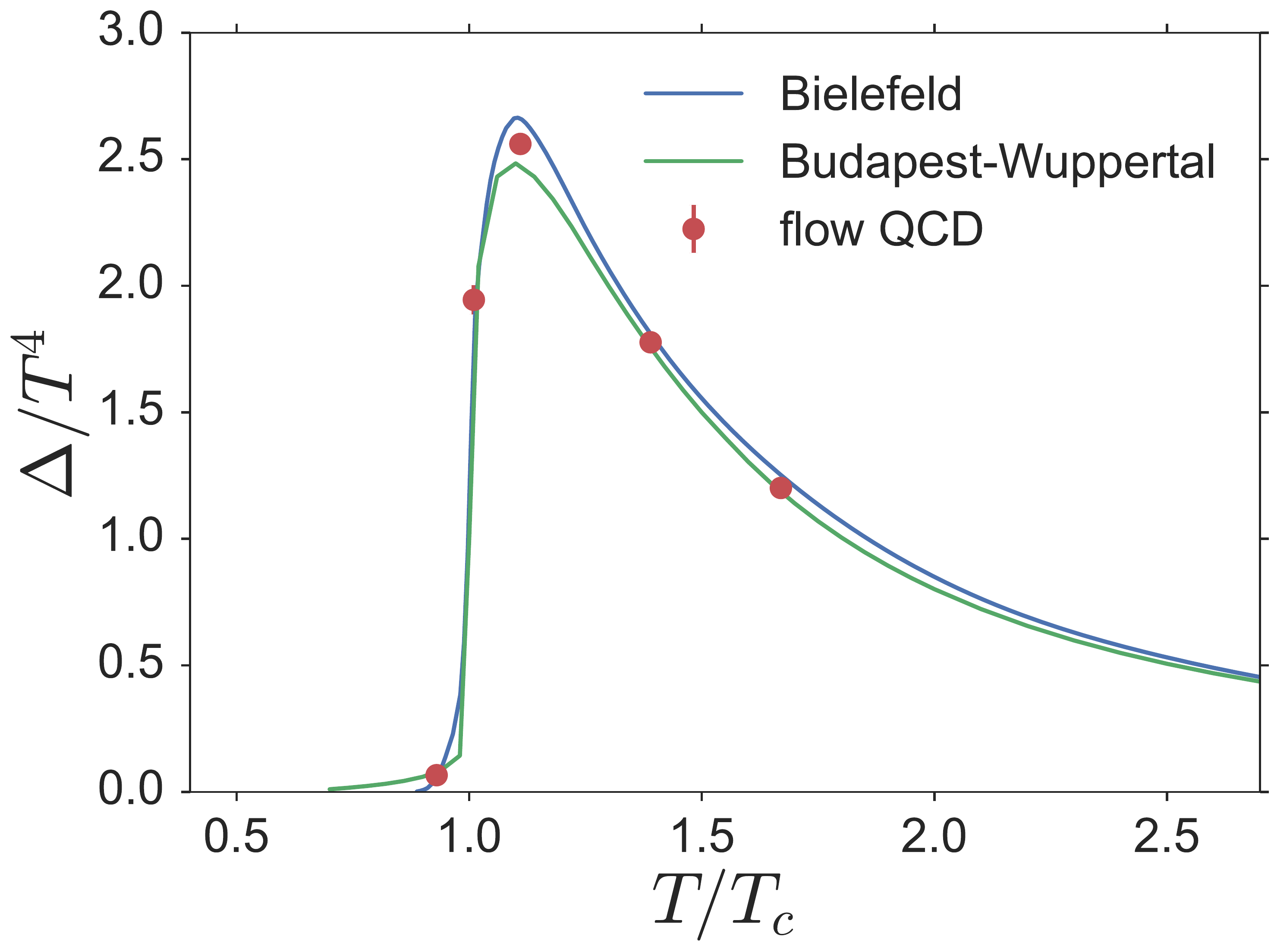}
 \includegraphics[width=0.49\textwidth,clip]{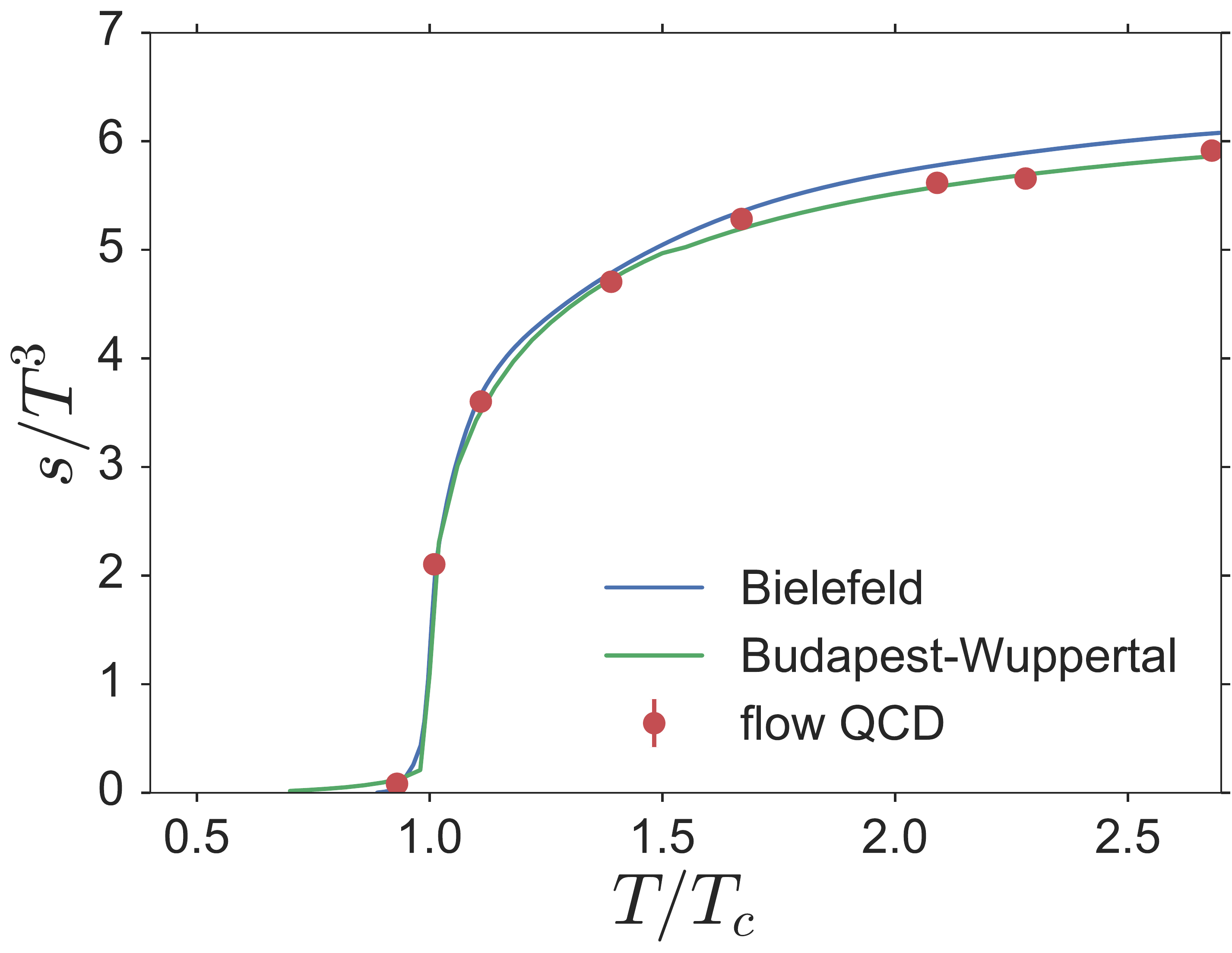}
 \caption{
   Temperature dependences of $\Delta/T^4$ and $s/T^3$ (red circles)
   in SU(3) YM theory~\cite{Kitazawa:2016dsl}
   together with the previous studies based on the integral method
   (solid lines)~\cite{Boyd:1996bx,Borsanyi:2012ve}.
   }
 \label{fig:therm}
\end{figure}

The $T$ dependence of thermodynamic quantities
$\Delta/T^4$ and $s/T^3$ in SU(3) YM theory obtained by this
step is shown in Fig.~\ref{fig:therm}
by the red circles~\cite{Kitazawa:2016dsl}.
In the figure, the results obtained by the
conventional integral method~\cite{Boyd:1996bx,Borsanyi:2012ve}
are also plotted.
It is remarkable that the values of $\Delta/T^4$ and $s/T^3$ 
obtained by the completely different methods agree well with each other.
This agreement suggests that EMT is successfully analyzed 
in the lattice simulation with the gradient flow
by the procedure introduced in the previous section.

Recently, novel methods to measure thermodynamics
in lattice gauge theory have been proposed~\cite{Giusti:2015daa,
  Giusti:2016iqr,Caselle:2016wsw,Caselle:2018kap}
besides the integral and gradient flow methods,
and they are applied to SU(3) YM theory.
As summarized in Refs.~\cite{Caselle:2018kap,Iritani:2018idk},
all these results agree well with each other, but 
there exists small but statistically significant
discrepancy above but near $T_c$.
Understanding the origin of this difference is an important future
study in the accurate measurement of thermodynamics.

\begin{figure}[t]
  \centering
  \includegraphics[width=0.49\textwidth,clip]{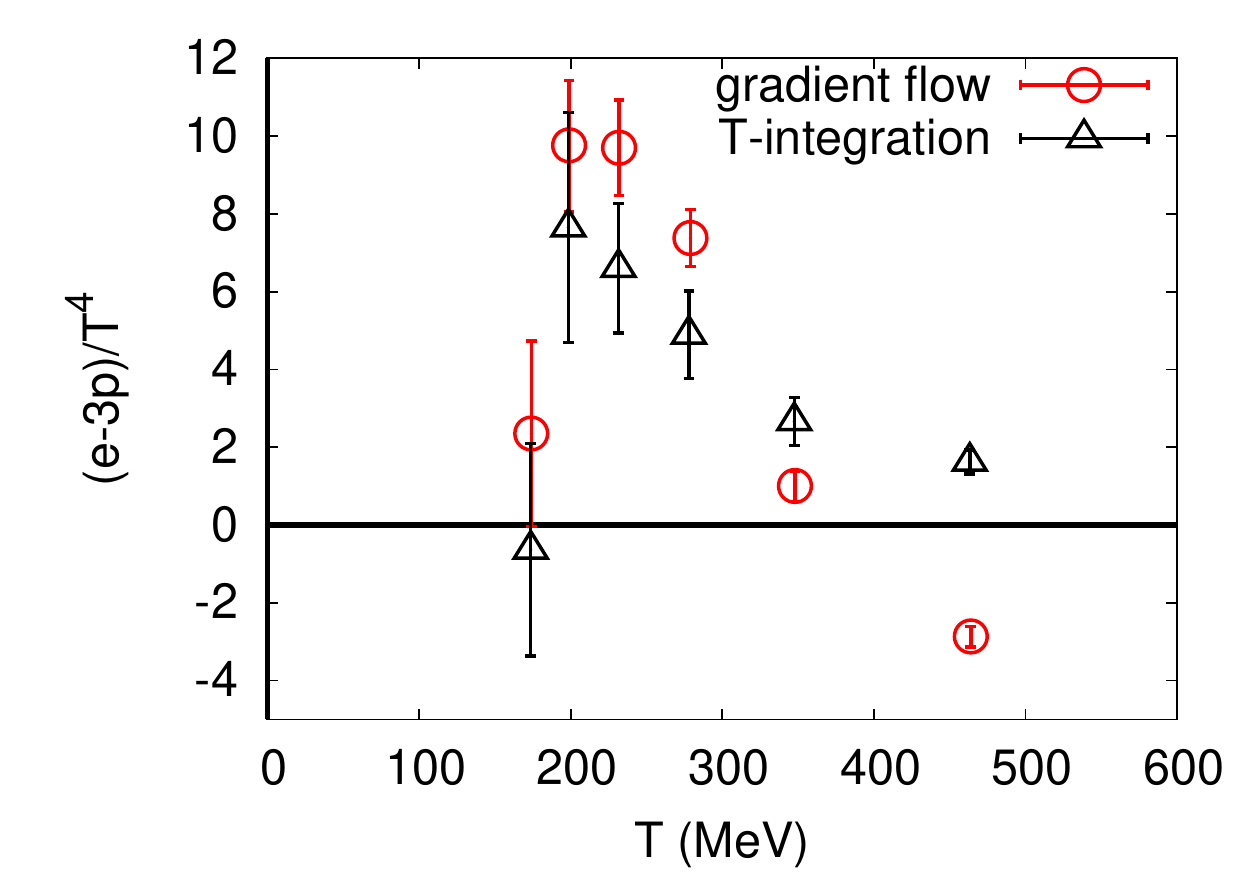}
  \includegraphics[width=0.49\textwidth,clip]{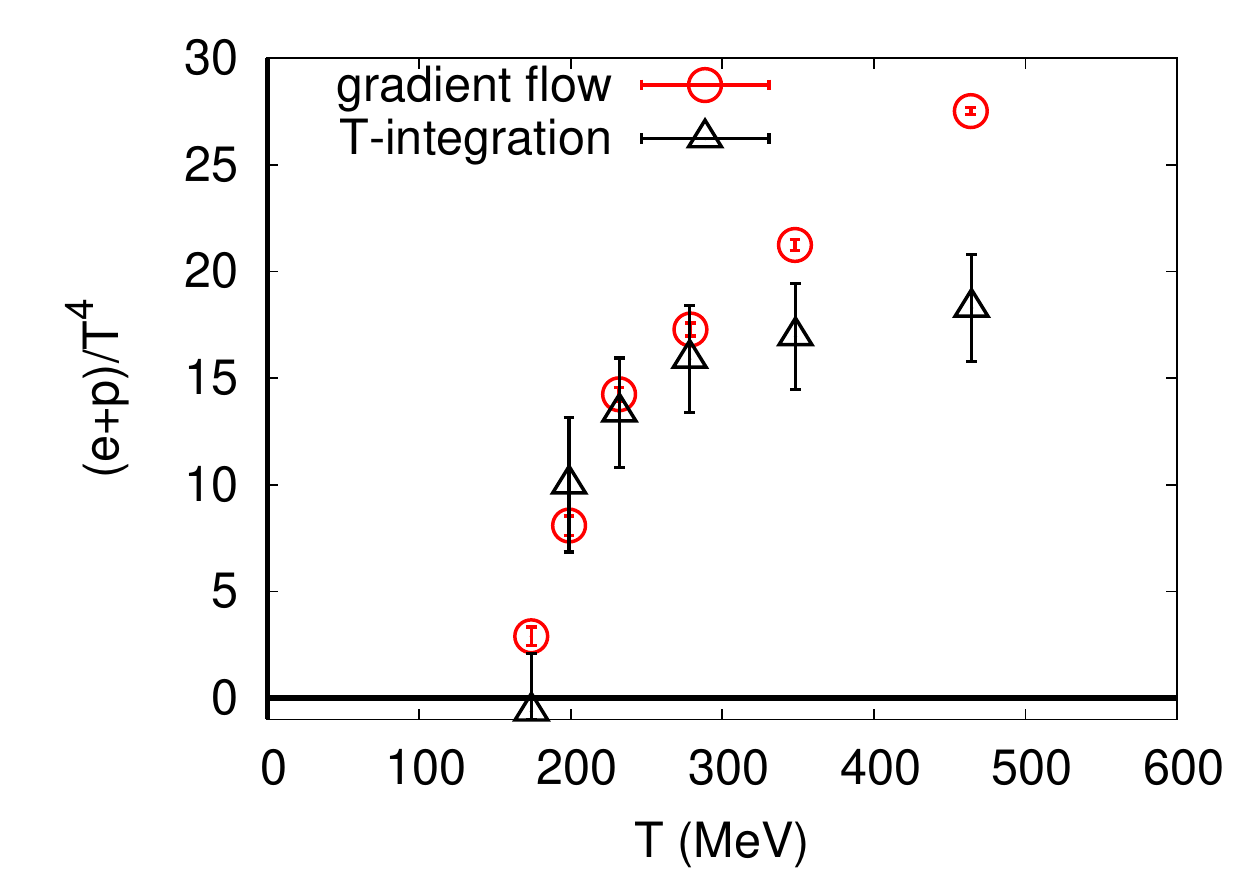}
  \caption{
    Temperature dependences of $\Delta/T^4$ and $s/T^3$ (red circles)
    in the (2+1)-flavor QCD together with the results obtained 
    by the integral method (black triangles)~\cite{Taniguchi:2016ofw}.
  }
 \label{fig:full-therm}
\end{figure}

The analysis of thermodynamics by the gradient flow method
can be applied to full QCD simulation with
fermions~\cite{Makino:2014taa,Taniguchi:2016ofw}.
In Fig.~\ref{fig:full-therm}, we show 
the $T$ dependences of $\Delta/T^4$ and $s/T^3$ in (2+1)-flavor QCD
obtained by the gradient flow method by the red circles, 
together with the results of the integral method
obtained on the same gauge configurations~\cite{Taniguchi:2016ofw}.
The mass of $u,d$ quarks is slightly heavy in this simulation;
$m_\pi/m_\rho\simeq0.63$.
Since the numerical simulation in this study 
is performed only for a single lattice spacing, 
the $t\to0$ extrapolation is taken without the continuum extrapolation
to obtain the final result in Fig.~\ref{fig:full-therm}.
Various extrapolating functions are adopted to take
the $t\to0$ extrapolation from numerical results at $\sqrt{2t}\gtrsim a$.
Fig.~\ref{fig:full-therm} shows that the two results 
agree well with each other except for the high temperature region
at which the $t\to0$ extrapolation is unstable.

\section{EMT correlation functions}
\label{sec:cor}

Next, we apply the EMT operator Eq.~(\ref{eq:T^R}) to the analysis of
the imaginary-time correlation function of EMT~\cite{Kitazawa:2017qab}
\begin{align}
  C_{\mu\nu;\rho\sigma}(\tau)
  = \frac1{T^5}\int_V d^3 x
  \langle T_{\mu\nu}(\vec{x},t) T_{\rho\sigma}(\vec{0},0) \rangle.
  \label{eq:C}
\end{align}
The EMT correlation function $C_{\mu\nu;\rho\sigma}(\tau)$ at nonzero
temperature contains various important information.
For example, the spatial components $C_{ij;kl}(\tau)$ are related to
transport coefficient through Kubo formula~\cite{Meyer:2011gj}.
However, $C_{\mu\nu;\rho\sigma}(\tau)$ is known to be extremely
noisy in lattice simulations~\cite{Meyer:2011gj}.

Here, as a first analysis of $C_{\mu\nu;\rho\sigma}(\tau)$ with 
the gradient flow method, we focus on the channels including
conserved quantities, i.e. $C_{44;44}(\tau)$, $C_{44;11}(\tau)$,
and $C_{41;41}(\tau)$.
Because of the energy and momentum conservation,
these correlators do not have a $\tau$ dependence for $\tau\ne0$.
Moreover, from thermodynamic relations they are given by
\begin{align}
C_{44;44}(\tau) = \frac{c_V}{T^3} , \qquad
C_{44;11}(\tau) = C_{41;41}(\tau) = -\frac{s}{T^3},
\label{eq:C_4}
\end{align}
for $\tau\ne0$ where $c_V$ is the specific heat per unit
volume.

\begin{figure}[t]
 \centering
 \includegraphics[width=0.32\textwidth,clip]{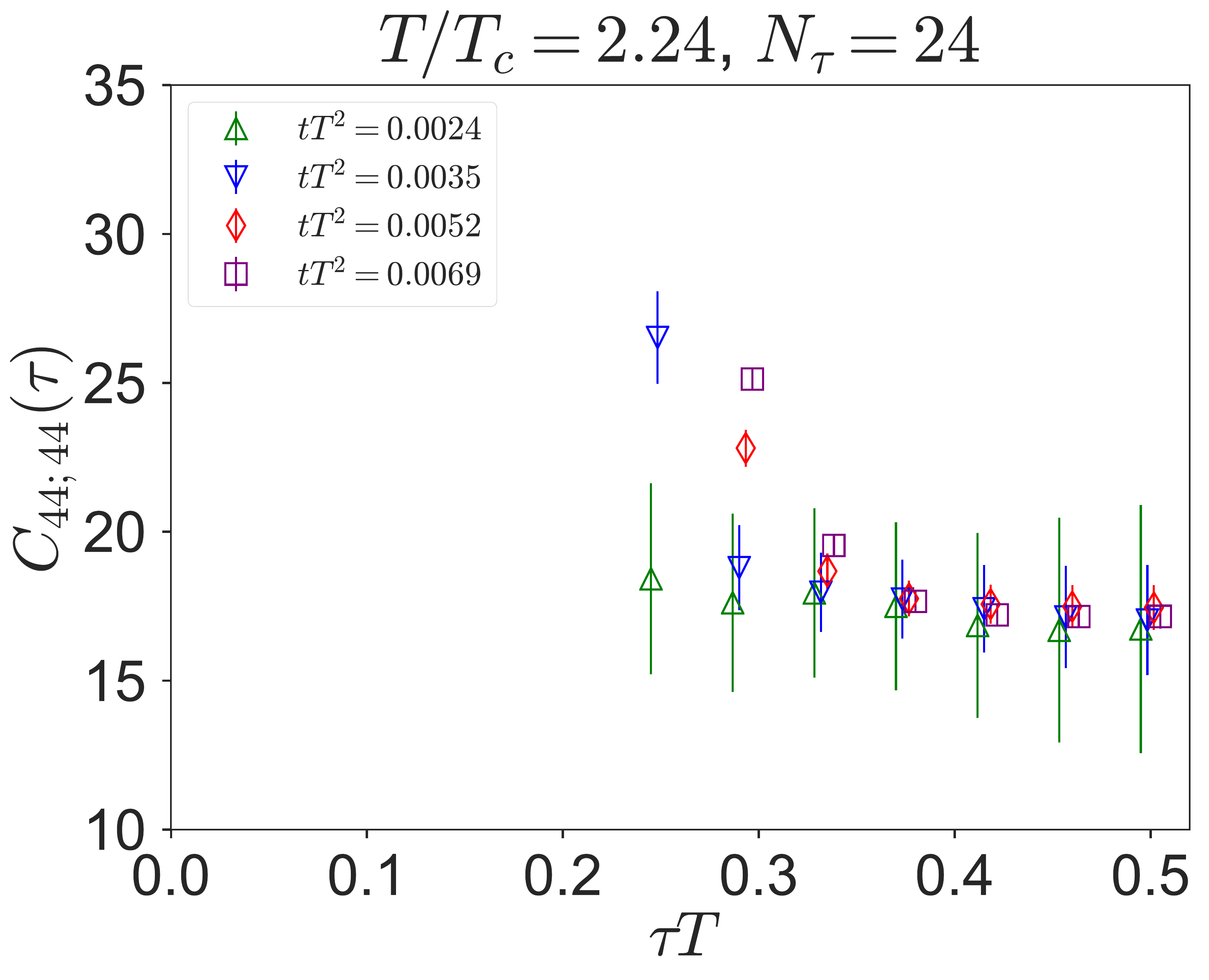}
 \includegraphics[width=0.32\textwidth,clip]{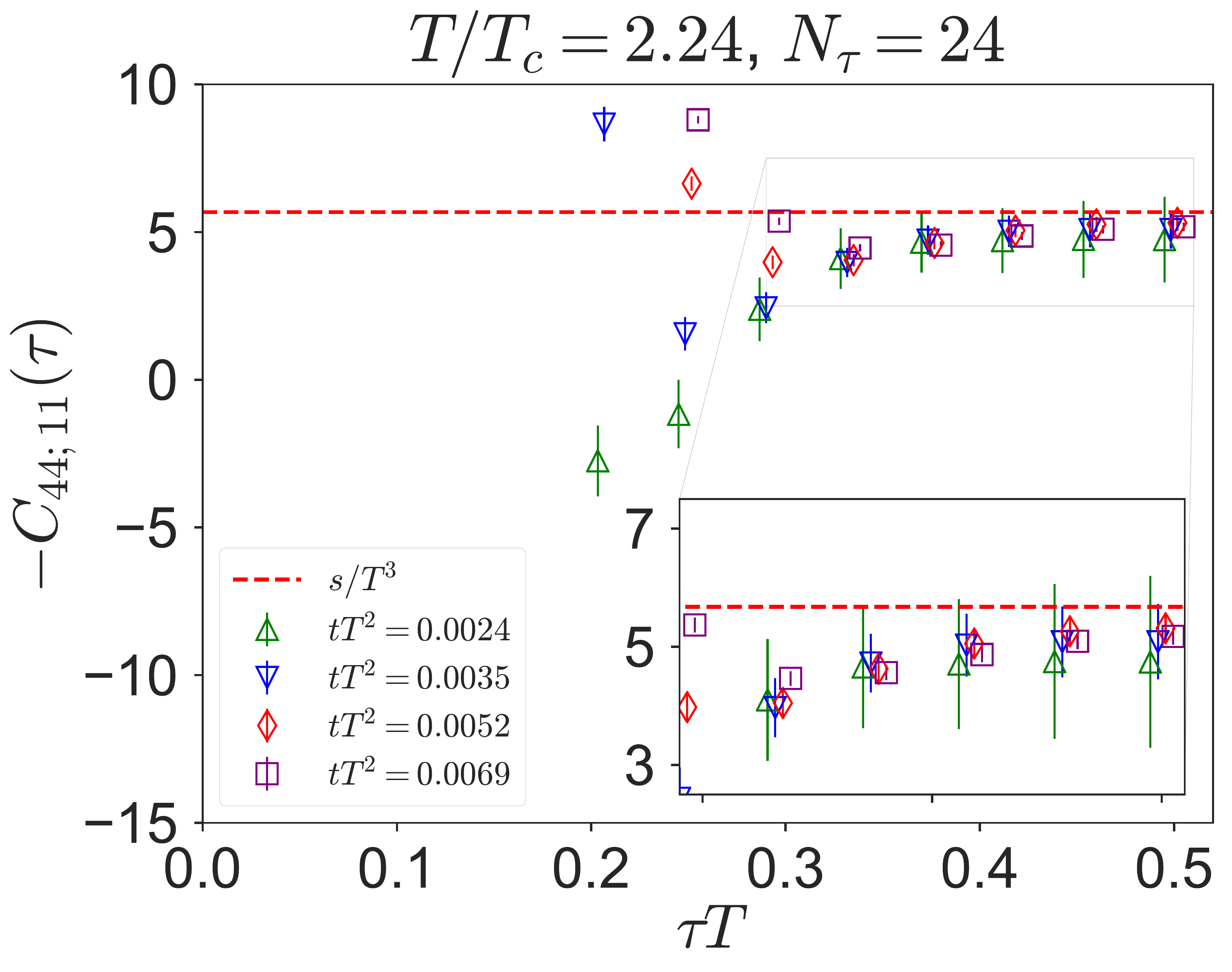}
 \includegraphics[width=0.32\textwidth,clip]{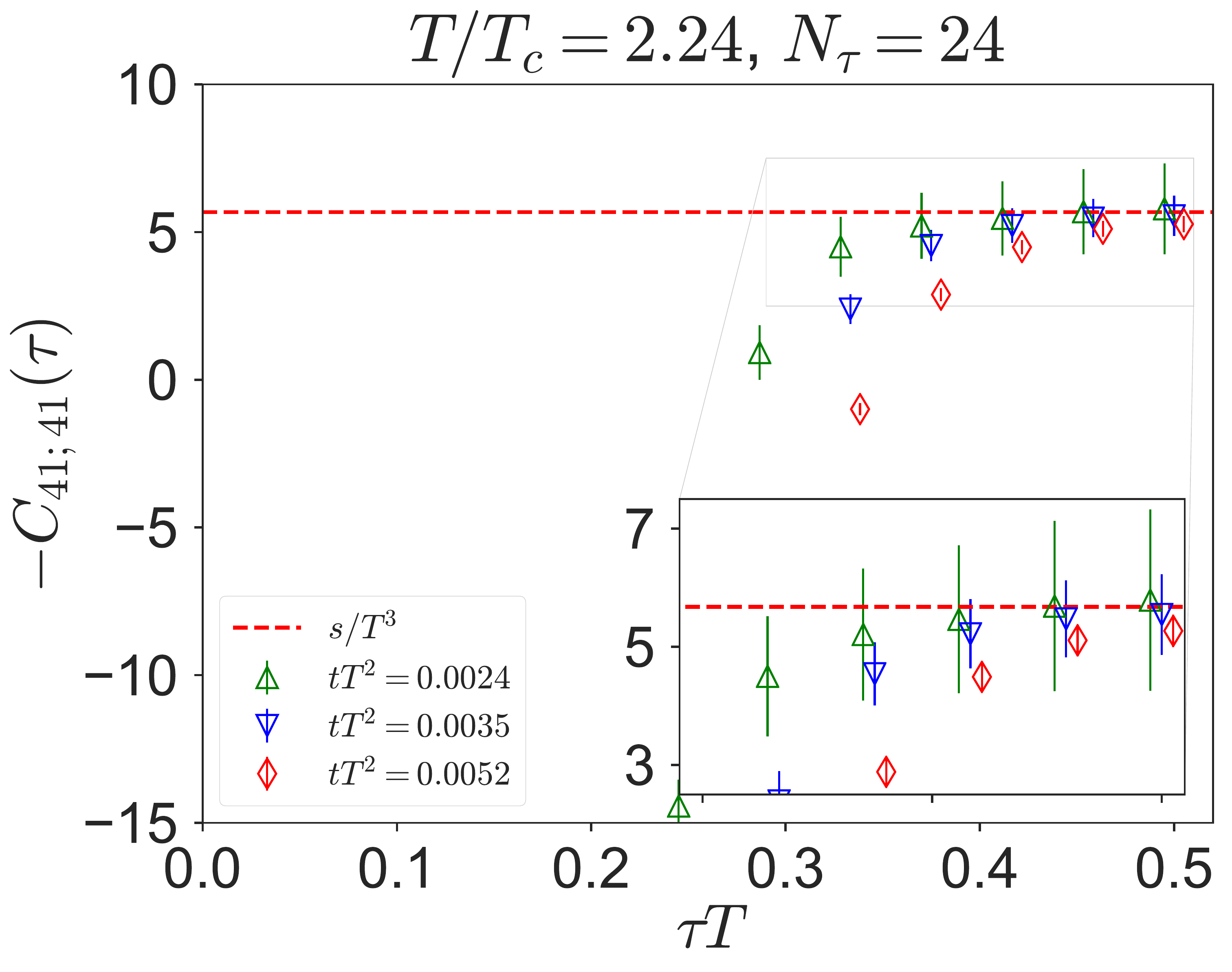}
 \caption{
   Correlation functions $C_{44;44}(\tau)$ (left),
   $C_{44;11}(\tau)$ (middle), and $C_{41,41}(\tau)$ (right)
   for several values of flow time $t$
   for $N_\tau=24$ and $T/T_c=2.24$.
   The red dashed lines on the middle and right panels show
   $s/T^3=(\varepsilon+p)/T^4$ obtained from
   the one-point function of the EMT with the same gauge
   configurations~\cite{Kitazawa:2017qab}.
 }
 \label{fig:cor}
\end{figure}

Shown in Fig.~\ref{fig:cor} are the correlation functions 
$C_{44;44}(\tau)$, $C_{44;11}(\tau)$, and $C_{41;41}(\tau)$
in SU(3) YM theory calculated on the lattice with $N_\tau=24$ and
$T/T_c=2.24$~\cite{Kitazawa:2017qab}.
The results are shown for several values of the flow time $t$.
The figure shows that there exists a $\tau$ independent
plateau for $\tau\gtrsim\sqrt{2t}$ which is consistent with
the conservation of energy and momentum.
The disappearance of the plateau at $\tau\lesssim\sqrt{2t}$
comes from the over-smearing due to the gradient flow.
To check the second equation in Eq.~(\ref{eq:C_4}),
in the middle and right panels of Fig.~\ref{fig:cor}
the value of $s/T^3$ is shown by the red-dashed line.
The panels show that $C_{44;11}(\tau)$ and $C_{41;41}(\tau)$
satisfy Eq.~(\ref{eq:C_4}).
More detailed verification of Eq.~(\ref{eq:C_4}) with the double
extrapolation $(t,a)\to(0,0)$, as well as the analysis of $c_V$
with the use of the first equation in Eq.~(\ref{eq:C}),
is carried out in Ref.~\cite{Kitazawa:2017qab}.

From these results, one finds that the EMT operator Eq.~(\ref{eq:T^R})
is successfully applied to the analysis of the EMT correlator Eq.~(\ref{eq:C}).
Therefore, it is an interesting subject to apply this method to 
the analysis of the spatial channels of Eq.~(\ref{eq:C})
which are relevant for the transport coefficient.
Recently, the analysis of these channels with the conventional EMT operator
in SU(3) YM theory
has been updated in Refs.~\cite{Pasztor:2018yae,Astrakhantsev:2018oue}.
Because these studies use multi-level algorithm, however,
it is difficult to extend the analysis to full QCD.
The analysis of Eq.~(\ref{eq:C}) in full QCD with the gradient flow
method is reported in Ref.~\cite{Taniguchi:2019eid}.

\section{Stress tensor distribution around $Q\bar{Q}$}
\label{sec:qqb}

Now we consider the spatial component of EMT
which is related to the stress tensor $\sigma_{ij}$ as in
Eq.~(\ref{eq:epsigma}).
In this section, we apply the EMT operator Eq.~(\ref{eq:T^R}) to
the analysis of the stress-tensor distribution
in static quark--anti-quark ($Q\bar{Q}$) systems in SU(3) YM
theory~\cite{Yanagihara:2018qqg}
in which the YM field strength is squeezed into
a quasi-one-dimensional flux-tube structure~\cite{Bali:2000gf}.
In the previous studies, the spatial structure of the flux tube 
has been investigated using 
the action density and the color electric field.
Compared with these observables, 
$\sigma_{ij}$ has clear physical meanings;
the stress tensor represents the local interaction
mediated by the distortion of the YM field.
Moreover, the stress tensor enables us to study this system in a
manifestly gauge invariant manner.

To prepare a static $Q\bar{Q}$ system on the lattice,
we use the standard Wilson loop $W(R,T)$
with static color charges at
$\vec{R}_{\pm}=(0,0, \pm  R/2)$ and in the temporal interval $[-T/2, T/2]$.
Then the expectation value of $T_{\mu\nu}(t,x)$
around the $Q\bar{Q}$ is obtained by 
\begin{align}
  \langle T_{\mu\nu}(t,x)\rangle_{Q\bar{Q}} =
  \lim_{T\to\infty} \frac{\langle T_{\mu\nu}(t,x) W(R,T)\rangle_0}
      {\langle W(R,T)\rangle_0} ,
      \label{eq:<T>_W} 
\end{align}
where $T \rightarrow \infty$ is to pick up the ground state of
$Q\bar{Q}$.
In actual numerical simulations, we use the APE smearing for each spatial
link to enhance the coupling of $W(R,T)$ to the $Q\bar{Q}$ ground state
with fixed $T$.
We also adopt the standard multi-hit procedure by replacing each temporal
link by its mean-field value to reduce the statistical noise.
The measurements of $T_{\mu\nu}(t,x)$ for different values of $t$
are made at the mid temporal plane $x_{\mu} =(\vec{x}, x_4=0)$,
while $W(R,T)$ is defined at $t=0$.

\begin{figure}[t]
 \centering
 \includegraphics[width=0.99\textwidth,clip]{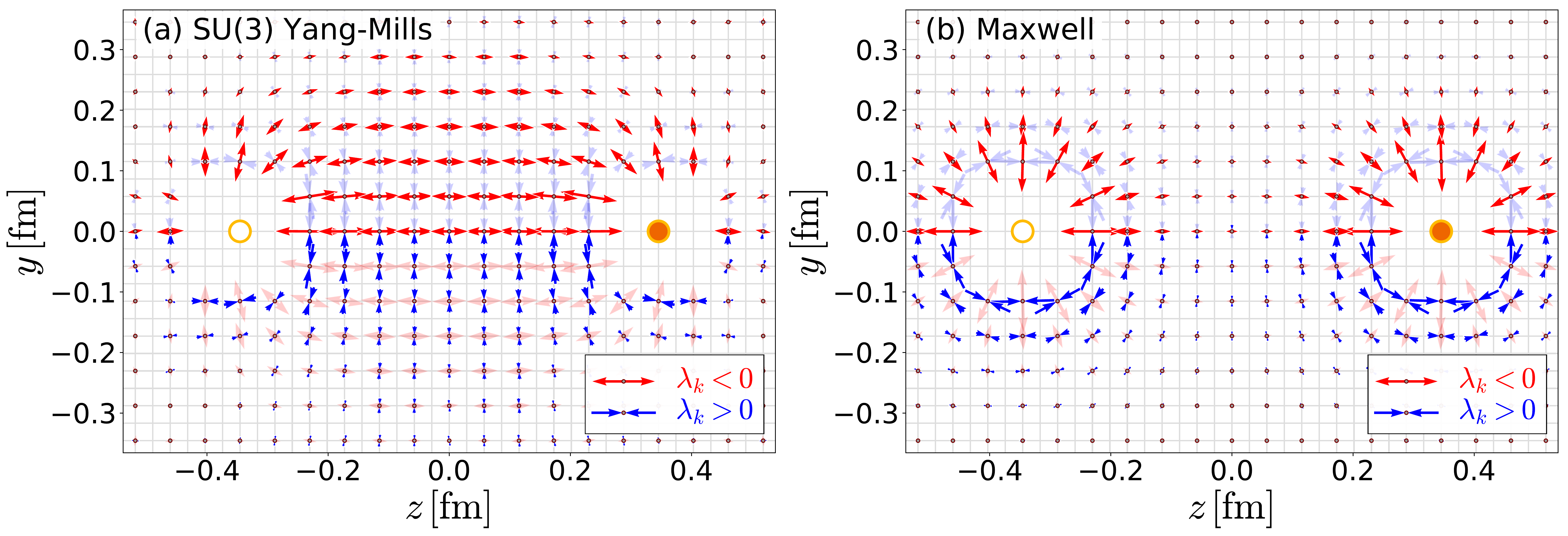}
 \caption{(a) Distribution of the principal axes of $T_{ij}$ for a
 $Q\bar{Q}$ system separated by $R=0.69$ fm in SU(3) Yang-Mills theory 
 with $a=0.029$~fm and $t/a^2=2.0$.
 (b) Distribution of the principal axes of the $T_{ij}$
 in classical electrodynamics between opposite charges.
 In both figures, the red (blue) arrows in the upper (lower) half plane are 
 highlighted~\cite{Yanagihara:2018qqg}. }
 \label{fig:stress-distribution}
\end{figure}

Before taking the double limit $(t,a)\to(0,0)$, we 
illustrate a qualitative feature of the distribution of $T_{ij}$
around  $Q\bar{Q}$ at fixed $a =0.029\ {\rm fm}$
and $t/{a}^2 = 2.0$ with $R=0.69\ {\rm fm}$.
In Fig.~\ref{fig:stress-distribution}~(a)~\cite{Yanagihara:2018qqg},
we show the two eigenvectors of the stress tensor defined by
\begin{align}
T_{ij}n_j^{(k)}=\lambda_k n_i^{(k)} \quad (k=1,2,3),
\label{eq:EV}
\end{align}
along with the principal axes of the local stress. 
The eigenvector with negative (positive) eigenvalue 
is  denoted by the red outward (blue inward) arrow
with its length proportional to $\sqrt{|\lambda_k |}$:
\begin{eqnarray}
 \red{\leftarrow} \hspace{-0.05cm} {\tiny{\circ}} \hspace{-0.05cm}  \red{\rightarrow} : \lambda_k<0, 
 \qquad  \blue{\rightarrow} \hspace{-0.05cm}  {\tiny{\circ}}  \hspace{-0.05cm}  \blue{\leftarrow}: \lambda_k>0.
\end{eqnarray}
Neighboring volume elements are pushing (pulling) with each other
along the direction of blue (red) arrow.
The spatial regions near $Q$ and $\bar{Q}$, which 
would suffer from over-smearing, are excluded in the figure.
Spatial structure of the flux tube is clearly 
revealed through the 
stress tensor in Fig.~\ref{fig:stress-distribution}~(a) in a gauge 
invariant way.
This is in contrast to the same plot of the principal axes of $T_{ij}$
for opposite charges in classical electrodynamics
shown in Fig.~\ref{fig:stress-distribution}~(b).

We note that the red arrows in Fig.~\ref{fig:stress-distribution} are
naturally interpreted as the direction of the line of field.
In this sense, Fig.~\ref{fig:stress-distribution}~(a) is a first
gauge invariant illustration of
{\it the line of the color electric field in YM theory}.

\begin{figure}[t]
  \centering
  \includegraphics[width=0.95\textwidth,clip]{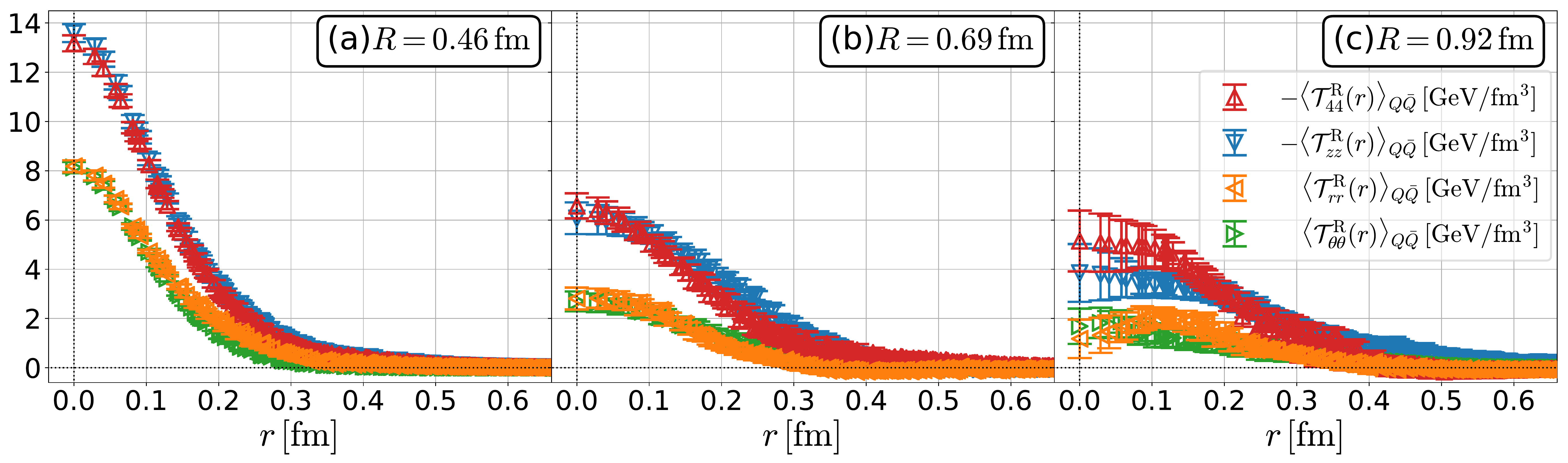}
  \caption{
EMT distribution on the mid-plane after the double limit
 $-\langle T^{\rm R} _{cc}(r) \rangle_{Q\bar{Q}} $
 and $-\langle T^{\rm R} _{44}(r) \rangle_{Q\bar{Q}} $
 in the cylindrical coordinate system for three different values of the $Q\bar{Q}$ distance $R$~\cite{Yanagihara:2018qqg}.
  \label{fig:mid}
  }
\end{figure}

Next, we focus on the mid-plane between the $Q\bar{Q}$
with $z=0$
and extract the stress-tensor distribution by taking the double
extrapolation $(t,a)\to(0,0)$~\cite{Yanagihara:2018qqg}.
On the mid-plane, it is convenient to use 
the cylindrical coordinate system $c=(r,\theta, z)$ with
$r=\sqrt{x^2+y^2}$ and $0 \le \theta < 2 \pi$.
One can show that the EMT on the mid-plane is diagonalized
in this coordinate as
\begin{align}
T_{c c'}^R(x) =
{\rm diag} ( T_{rr}^R (r) , T_{\theta\theta}^R (r) , T_{zz}^R (r) ).
\label{eq:diag}
\end{align}

Shown in Fig.~\ref{fig:mid} is the $r$ dependence of the resulting EMT,
i.e. the stress tensor  $ - \langle T^{\rm R}_{cc}(r) \rangle_{Q\bar{Q}} $
and the
energy density  $-\langle T^{\rm R}_{44}(r)\rangle_{Q\bar{Q}} $
with three $Q\bar{Q}$ distances $R=0.46, 0.69, 0.92$~fm~\cite{Yanagihara:2018qqg}.
From the figure, one finds several notable features.
First,
approximate degeneracy 
$ \langle T^{\rm R}_{44}(r)  \rangle_{Q\bar{Q}}  \simeq
\langle T^{\rm R}_{zz}(r) \rangle_{Q\bar{Q}}$
as well as $ \langle T^{\rm R}_{rr}(r)  \rangle_{Q\bar{Q}}  \simeq
\langle T^{\rm R}_{\theta \theta}(r)  \rangle_{Q\bar{Q}} $
is found for a wide range of $r$ and $R$.
Second, 
the nonzero value of the trace of EMT
$ \langle T^{\rm R}_{\mu \mu}(r) \rangle_{Q\bar{Q}} =
\langle T^{\rm R}_{44}(r) +  T^{\rm R}_{zz}(r) +
T^{\rm R}_{rr}(r)  +  T^{\rm R}_{\theta \theta}(r) \rangle_{Q\bar{Q}} <  0 $
is observed, which suggests
the partial restoration of the scale symmetry broken in the YM vacuum.
Finally, the radius of the flux tube, typically about $0.2$~fm, 
seems to become wider with increasing $R$.

\begin{figure}
  \centering
  \includegraphics[width=0.47\textwidth,clip]{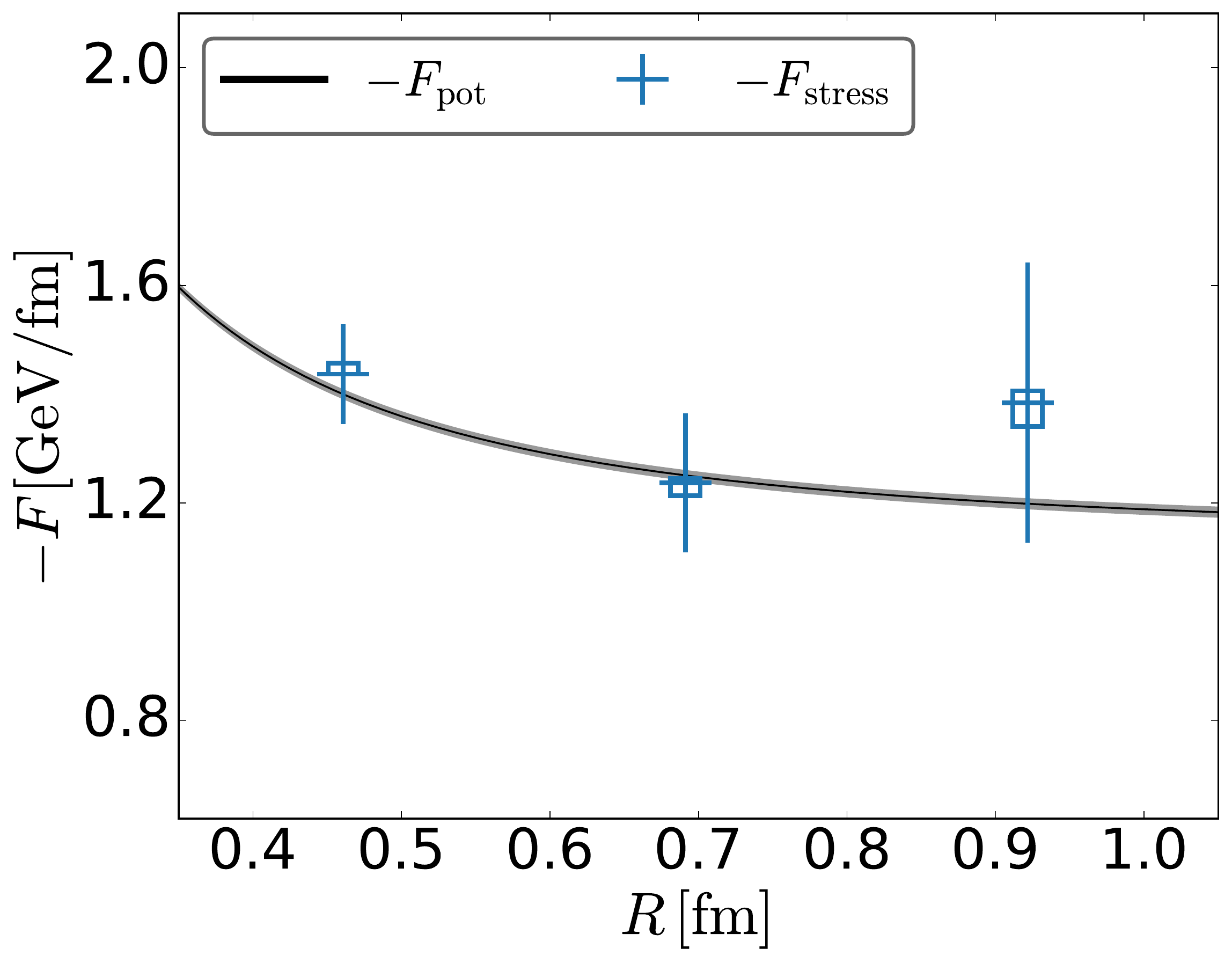}
  \caption{
  $R$ dependence of the $Q\bar{Q}$ forces,
   $-F_{\rm stress}$ and $-F_{\rm pot}$, obtained by the Wilson loop
 and the stress tensor, respectively.
 Error bars and rectangular boxes for the latter represent the  statistical and systematic errors,
 respectively~\cite{Yanagihara:2018qqg}.
 \label{fig:R}
  }
\end{figure}

Next, we consider a non-trivial consistency check for the uniqueness
of the force~\cite{Yanagihara:2018qqg}.
The force acting on the charge located at $z>0$ can be obtained 
by two different manners; 
(i) through the $Q\bar{Q}$ potential $V(R)$ as
\begin{align}
F_{\rm pot}=-dV(R)/dR
\end{align}
and (ii) from the surface integral of the stress-tensor 
surrounding the charge,
\begin{align}
F_{\rm stress} = - \int  \langle T_{zj}(x) \rangle_{Q\bar{Q}}\  dS_j .
\end{align}
For $F_{\rm pot}$,
we use the numerical data of $V(R)$ obtained from the Wilson loop
at $a=0.038$~fm.
For $F_{\rm stress}$, we take the mid-plane for the surface integral:
$F_{\rm stress} = 2 \pi \int_0^{\infty} \langle T_{zz}(r)
\rangle_{Q\bar{Q}}  \ r dr $.
In Fig.~\ref{fig:R}, $- F_{\rm pot}$ and $- F_{\rm stress}$
thus obtained are shown by the solid line and the
horizontal bars, respectively~\cite{Yanagihara:2018qqg}.
The figure shows the agreement between the two quantities within the errors,
which is a first numerical evidence that the ``action-at-a-distance''
$Q\bar{Q}$ force can be described
by the local properties of the stress tensor in YM theory.

\section{Analysis of $Q\bar{Q}$ system in Abelian-Higgs model}
\label{sec:AH}

Finally, let us investigate the behavior of $T_{\mu\nu}(r)$
in Fig.~\ref{fig:mid} in more detail especially focusing on 
the approximate degeneracy and separation of each channel, 
$T_{rr}(r) \simeq T_{\theta\theta}(r) < T_{zz}(r)$~\cite{prep}.

First, from the momentum conservation, $\partial_i T_{ij}=0$,
one can show that the EMT
in the cylindrical coordinates Eq.~(\ref{eq:diag}) satisfies
$ \partial_r(rT_{rr})
 -T_{\theta\theta}
 +r \partial_z T_{rz}=0.
$
Then, by further assuming that the flux tube is sufficiently long so that 
it has a translational invariance along the $z$ direction,
the $z$ derivative vanishes and one obtains,
\begin{align}
 \partial_r(rT_{rr}) =T_{\theta\theta}.
 \label{eq:cons}
\end{align}
From this differential equation it is concluded that 
$T_{rr}(r)$ and $T_{\theta\theta}(r)$ do not degenerate
except for the case $T_{rr}(r)=T_{\theta\theta}(r)=0$.
Moreover, by integrating out both sides of Eq.~(\ref{eq:cons}) by $r$ and
using the boundary conditions $rT_{rr}(r)\to0$ for $r\to0$ and $r\to\infty$,
one obtains
\begin{align}
 \int_0^\infty dr T_{\theta\theta}(r) =0,
\end{align}
which means that $T_{\theta\theta}(r)$ must change the sign at least once.
Such behaviors of $T_{rr}(r)$ and $T_{\theta\theta}(r)$, however,
are not observed in Fig.~\ref{fig:mid}
even at the largest $Q\bar{Q}$ distance.
It is therefore suggested that the finite-length effect of the flux tube
is not negligible even at $R=0.92$~fm in SU(3) YM theory.

Next, in order to get ideas on physics behind the results 
in Fig.~\ref{fig:mid}
we study the EMT distribution in a specific model.
For this purpose, here we employ the Abelian-Higgs (AH) model
\begin{align}
 \mathcal{L}_{\mathrm{AH}}
 %&=-\frac{1}{4g^2}(\partial_\mu B_\nu(x)-\partial_\nu B_\mu(x))^2
 %+|(\partial_\mu +iB_\mu(x))\chi(x)|^2-\lambda(|\chi(x)|^2-v^2)^2
 %\\
 =-\frac{1}{4g^2}F_{\mu\nu}^2
 +|D_\mu\chi(x)|^2-\lambda(|\chi(x)|^2-v^2)^2,
 \label{eq:AH}
\end{align}
which is the relativistic extension of the Ginzburg-Landau model,
with the field strength
$F_{\mu\nu}=\partial_\mu A_\nu(x)-\partial_\nu A_\mu(x)$ and 
the covariant derivative $D_\mu=\partial_\mu +iA_\mu(x)$.
The AH model has classical solutions with a static
magnetic vortex.
When this model is viewed as an effective model
of QCD according to the dual-superconductor picture~\cite{tHooft},
the vortex solution is considered as an analogue of the flux tube
in YM theory.
In the following, we thus study the EMT distribution around the classical
vortex solution of Eq.~(\ref{eq:AH}) with the winding number $n=1$.~\cite{prep}

\begin{figure}[t]
  \centering
  \includegraphics[width=0.32\textwidth,clip]{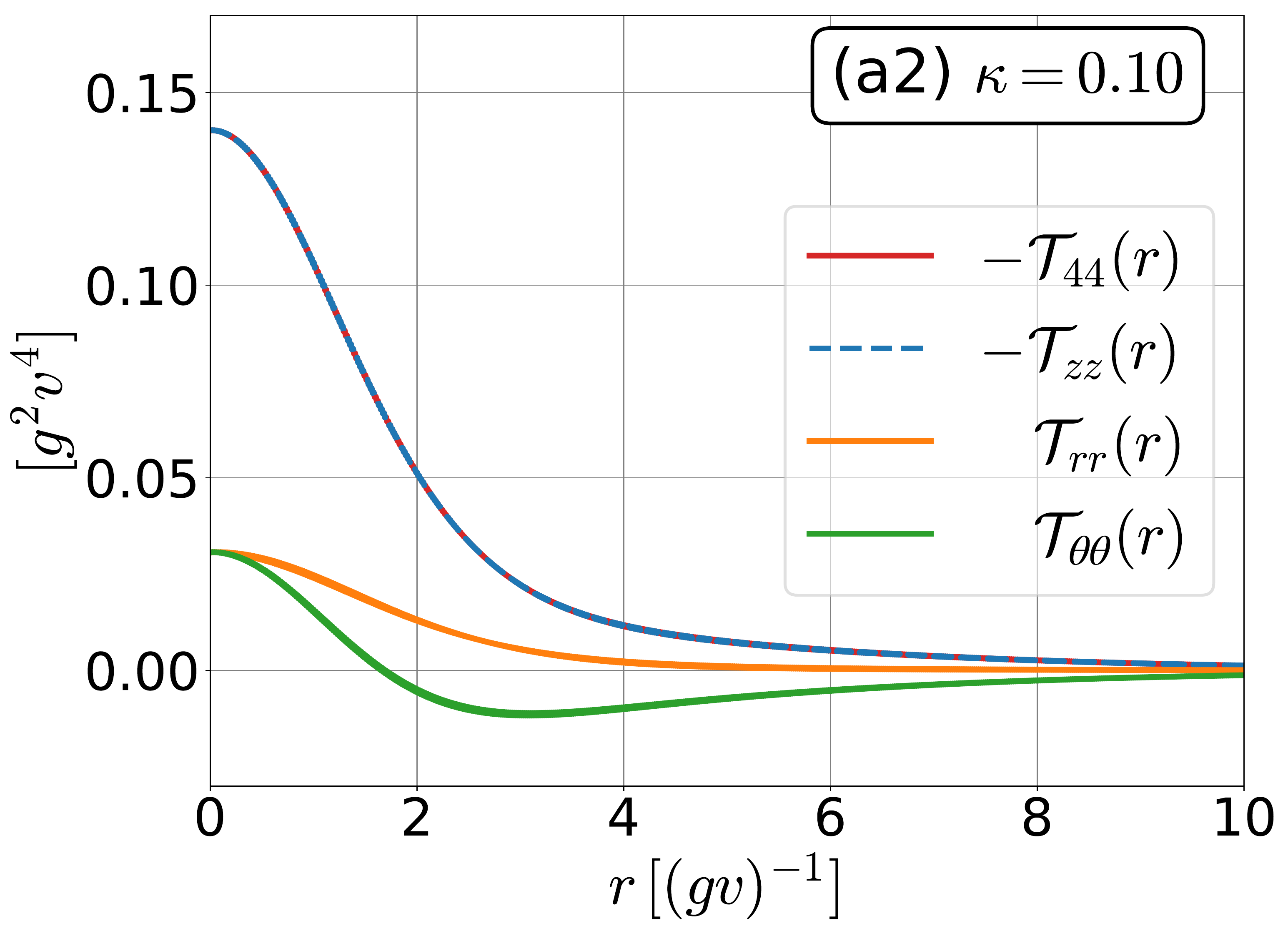}
  \includegraphics[width=0.32\textwidth,clip]{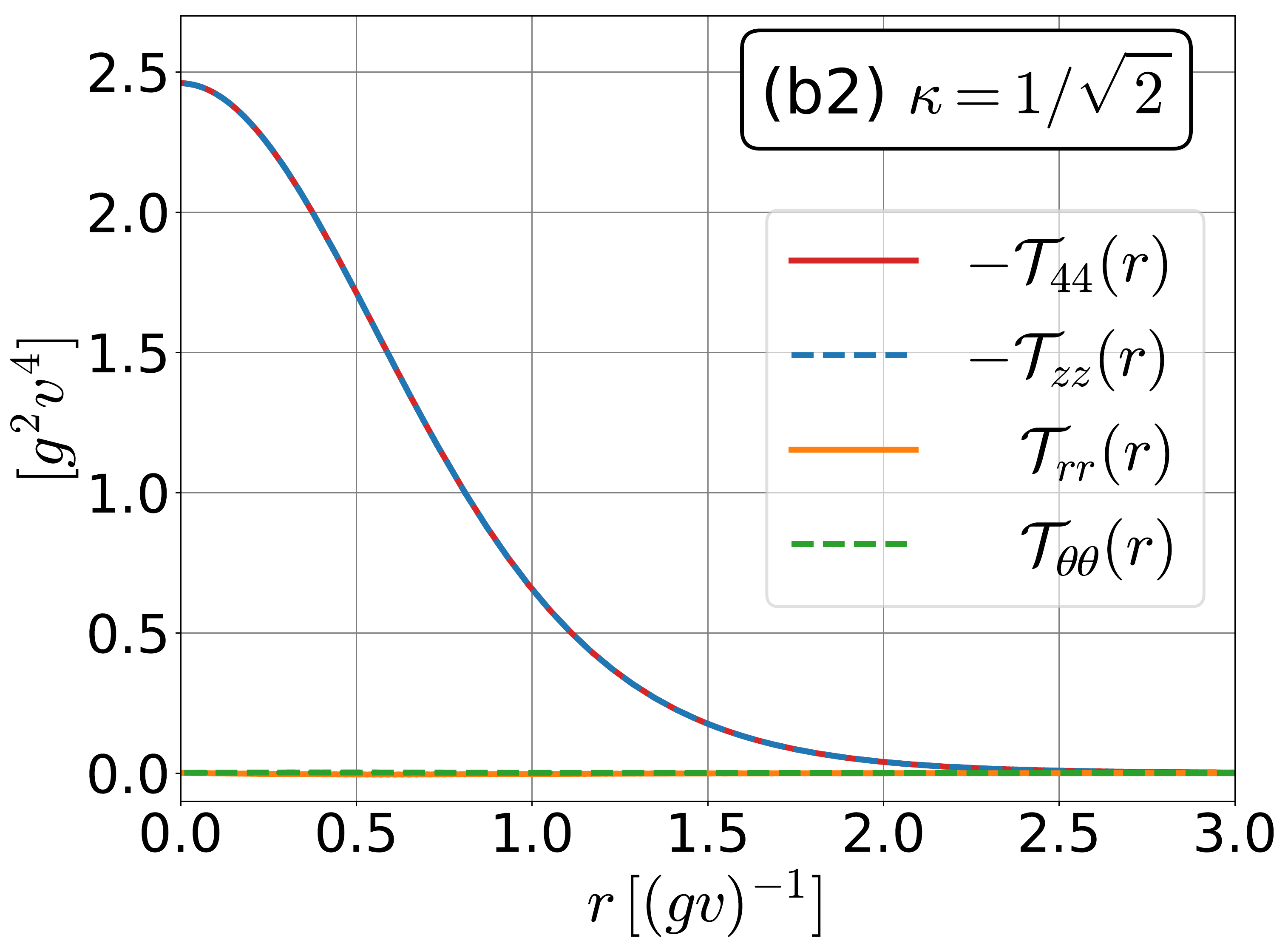}
  \includegraphics[width=0.32\textwidth,clip]{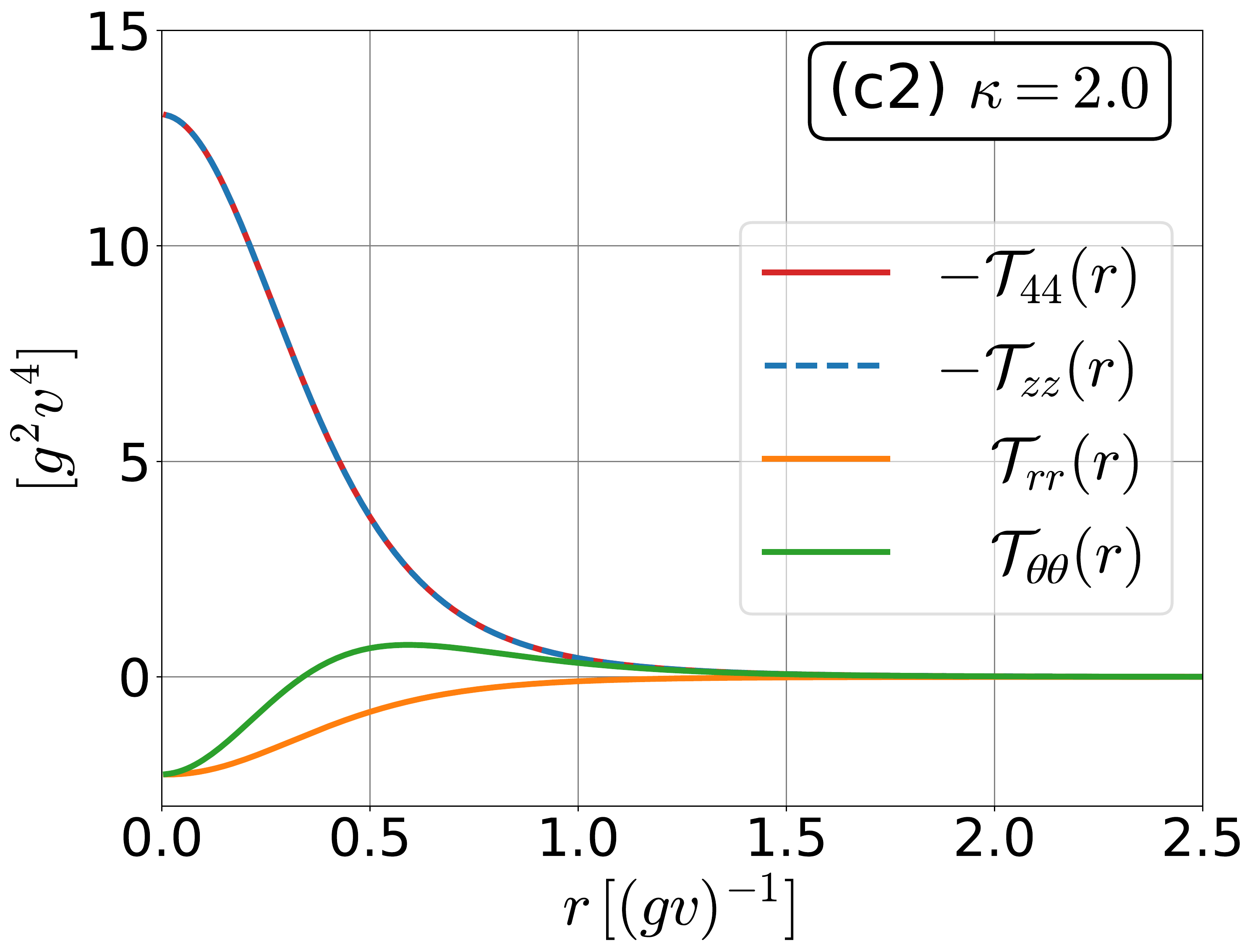}
  \caption{
    EMT distribution around the infinitely-long vortex in AH
    model~\cite{prep}.
  \label{fig:AH}
  }
\end{figure}

We first consider the EMT distribution around an infinitely-long and
straight vortex.
In Fig.~\ref{fig:AH}, we show the EMT in cylindrical
coordinates on the cross section of the vortex as a function of radius $r$.
Three panels show the results for different values of the
Ginzburg-Landau parameter $\kappa = \sqrt{\lambda}/g$.
From the figure, one finds that $T_{rr}(r)$ and $T_{\theta\theta}(r)$ have
a clear separation except with $\kappa = 1/\sqrt{2}$ at which 
$T_{rr}(r)=T_{\theta\theta}(r)=0$.
This separation is a model independent feature anticipated from 
the momentum conservation Eq.~(\ref{eq:cons}).

\begin{figure}[t]
  \centering
  \includegraphics[width=0.32\textwidth,clip]{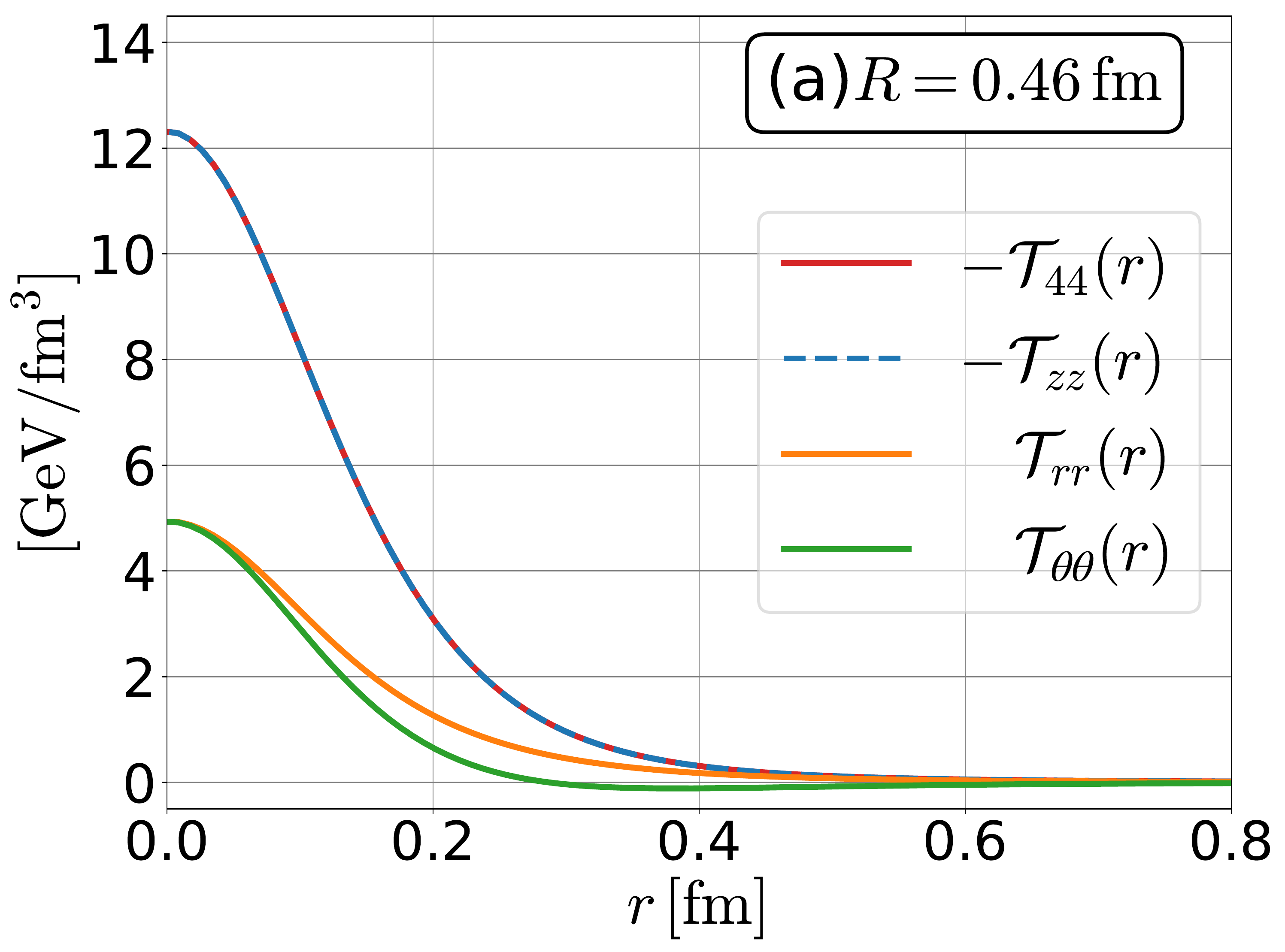}
  \includegraphics[width=0.32\textwidth,clip]{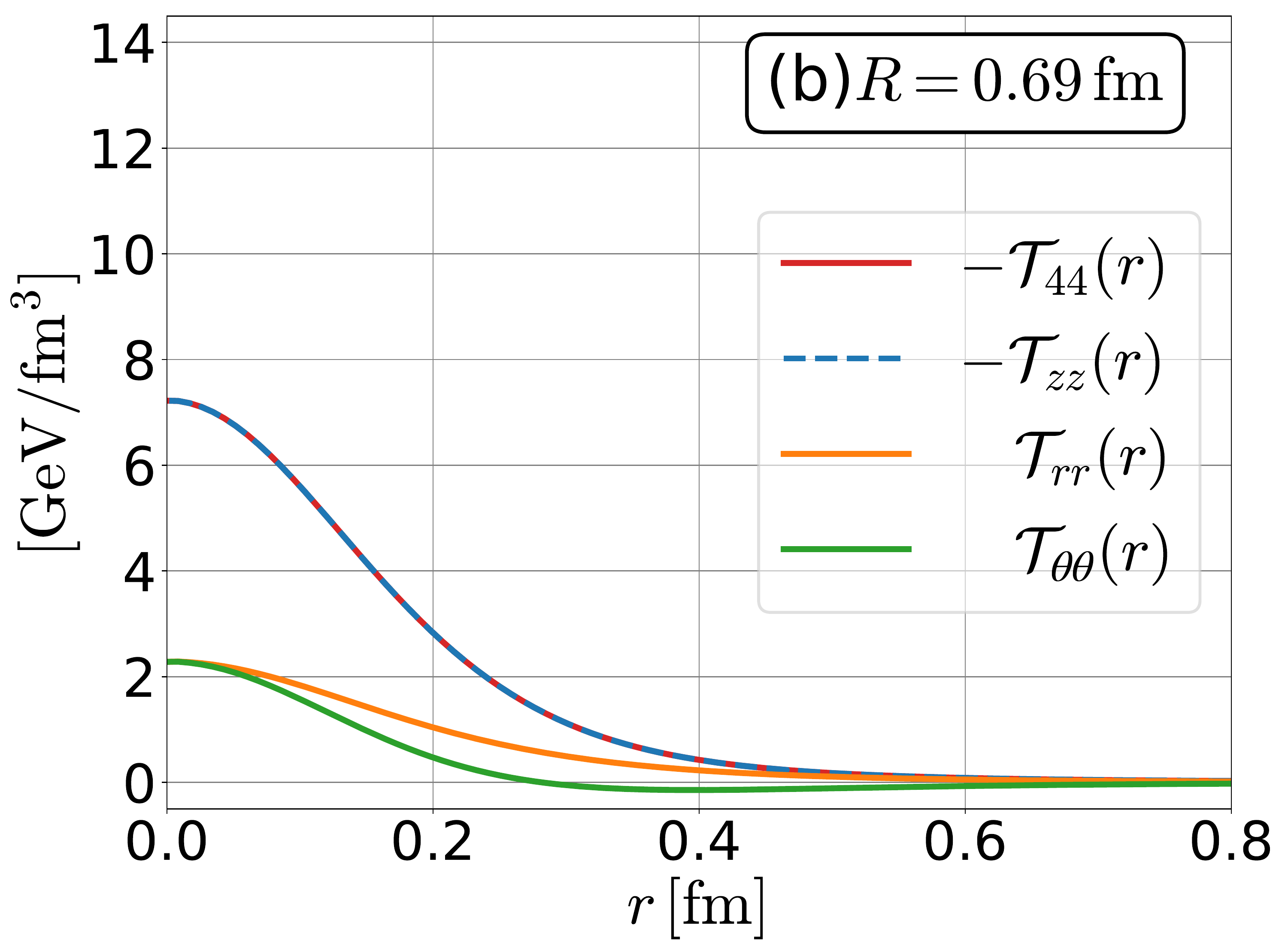}
  \includegraphics[width=0.32\textwidth,clip]{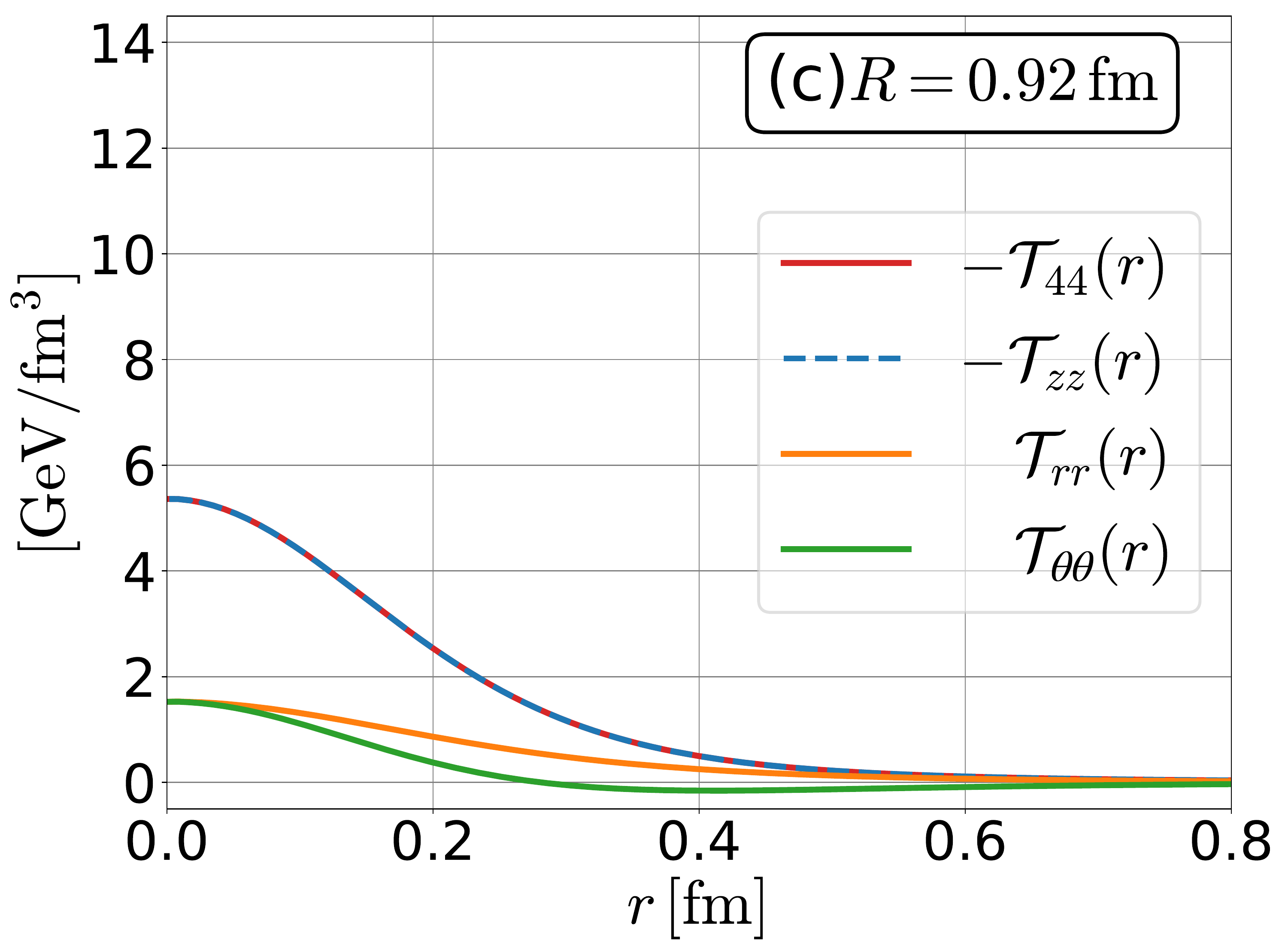}
  \caption{
    EMT distribution on the mid-plane between two magnetic monopoles
    for three different values of the distance
    between the monopoles $R$~\cite{prep}.
  \label{fig:AHfinite}
  }
\end{figure}

Next, we investigate the vortex with finite length $R$.
In AH model, the magnetic vortex with boundaries is obtained
by putting two magnetic monopoles with a unit charge but opposite
signs~\cite{Koma:2003gq}.
Shown in Fig.~\ref{fig:AHfinite} are examples of the EMT distribution
on the mid-plane between two monopoles~\cite{prep}.
The physical dimension in the figure is introduced by
setting the energy density per unit length of the infinitely-long
vortex to be the string tension obtained on the lattice.
The model parameters are chosen so that the values of 
$T_{cc}(r)$ at $r=0$ are consistent with the lattice result
in Fig.~\ref{fig:mid} after setting
the length $R$ to be equivalent with the $Q\bar{Q}$ distance in
the lattice simulations.
The figure shows that the difference between $T_{rr}(r)$ and
$T_{\theta\theta}(r)$ becomes small compared to Fig.~\ref{fig:AH}.
However, $T_{rr}(r)$ and $T_{\theta\theta}(r)$ still have a separation
which would be inconsistent with Fig.~\ref{fig:mid}.
It is thus suggested that the degeneracy of these channels is a
unique feature of the flux tube in SU(3) YM theory which cannot be
reproduced by AH model.

\section{Summary and outlook}
\label{sec:summary}

In this proceeding, we reviewed recent attempts to perform the measurement
of EMT on the lattice with the gradient flow and its applications to 
thermodynamics, correlations, and the stress-tensor distribution inside
the flux tube.
All these results show that EMT is successfully analyzed in lattice
simulations with a reasonable statistics.

Now that the analysis of the EMT on the lattice is established,
there are many further applications, as EMT is one of the most
fundamental observables in physics.
For example, it is quite interesting to extend 
the study of the flux tube to nonzero temperature, multi-quark systems,
and excited states.
It is also an interesting future study to analyze the EMT distribution
inside hadrons~\cite{Burkert:2018bqq,Polyakov:2018zvc,Shanahan:2018nnv}
using the gradient flow method.

This work was supported by JSPS KAKENHI No.~JP17K05442.
Numerical simulation was carried out on IBM System Blue Gene Solution 
at KEK under its Large-Scale Simulation Program,
Reedbush-U at the University of Tokyo,
%under Initiative on Promotion of Supercomputing for Young or Women Researchers,
and OCTOPUS at Osaka University.

\end{document}